\documentclass[aps,prb,twocolumn,groupedaddress,showpacs]{revtex4-1}

\usepackage{amsmath}
\usepackage{amsfonts}
\usepackage{amssymb}
\usepackage{MnSymbol}
\usepackage{xspace}
\usepackage{wasysym}
\usepackage{graphicx}
\usepackage[utf8]{inputenc}
\usepackage{hyperref}
\usepackage{braket}
\usepackage{color}
\usepackage{dsfont}

\setcitestyle{square,numbers}

\DeclareMathOperator{\Tr}{Tr}

\newcommand{\ie}[0]{i.e.\@\xspace}
\newcommand{\eg}[0]{e.g.\@\xspace}
\newcommand{\etal}[0]{\textit{et al.}\@\xspace}

\newcommand{\ham}[1]{\hat{H}_{#1}}
\newcommand{\hamcl}[1]{{H}_{#1}}

\newcommand{\fan}[1]{\hat{c}^{\vphantom\dagger}_{#1}}
\newcommand{\fcr}[1]{\hat{c}^{\dagger}_{#1}}
\newcommand{\fden}[1]{\hat{n}_{#1}}
\newcommand{\fbond}[1]{\hat{B}_{#1}}

\newcommand{\Q}[1]{\hat{q}_{#1}}
\renewcommand{\P}[1]{\hat{p}_{#1}}

\newcommand{\q}[1]{q_{#1}}

\newcommand{\kF}{k_F}

\newcommand{\sech}{\mathrm{sech}}

\newcommand{\im}{\mathrm{i}}

\newcommand{\absolute}[1]{\left| #1 \right|}
\newcommand{\expv}[1]{\left\langle #1 \right\rangle}

\newcommand{\expvc}[1]{\left\llangle #1 \right\rrangle_{C}}

\begin{document}


\title{Thermodynamic and spectral properties of adiabatic Peierls chains}


\author{Manuel Weber}
\author{Fakher F. Assaad}
\author{Martin Hohenadler}
\affiliation{\mbox{Institut f\"ur Theoretische Physik und Astrophysik,
Universit\"at W\"urzburg, 97074 W\"urzburg, Germany}}


\date{\today}

\begin{abstract}
We present exact numerical results for the effects of thermal fluctuations on the
experimentally relevant thermodynamic and spectral properties of Peierls chains.
To this end, a combination of classical Monte Carlo sampling and exact
diagonalization is used to study adiabatic half-filled Holstein and
Su-Schrieffer-Heeger models. The classical nature of the lattice
displacements in combination with parallel tempering permit simulations on
large system sizes and a direct calculation of spectral functions in the frequency domain.
Most notably, the long-range order and the associated Peierls gap give rise
to a distinct low-temperature peak in the specific heat. The closing
of the gap and suppression of order by thermal fluctuations involves in-gap
excitations in the form of soliton-antisoliton pairs, and is also reflected
in the dynamic density and bond structure factors as well as in the optical
conductivity. We compare our data to the widely used mean-field
approximation, and highlight relations to symmetry-protected topological phases
and disorder problems.
\end{abstract}

\pacs{71.38.-k, 71.20.Rv, 65.40.Ba}

\maketitle

\section{Introduction}

Quasi-one-dimensional (1D) materials exhibit exciting phenomena such as
spin-charge separation due to electron-electron interaction (\eg, in TTF-TCNQ \cite{PhysRevLett.88.096402}) or insulating charge-density-wave states as a
result of electron-phonon coupling (\eg, in blue bronze
\cite{TRAVAGLINI1983289}). While the ground-state properties of 1D 
spin or electron models can often be fully understood with the help of
bosonization \cite{Voit94} and numerical methods, models with phonons
remain a challenge. The calculation of thermodynamic or nonequilibrium
properties is even harder and requires further methodological improvements. 
The role of electron-phonon coupling for the relaxation of charge-density-wave
systems after photo-induced phase transitions is currently of particular
interest \cite{Yonemitsu20081}. 

Because of the Peierls instability, a 1D metal with one electron per unit
cell can undergo a transition to a dimerized state with long-range charge order
and a gap at the Fermi level \cite{Frohlich54,Peierls55}. Depending on the form of
the coupling, the charge order is either on the sites or the bonds. Even
neglecting electron-electron interaction, the Peierls state is affected by
quantum fluctuations, soliton excitations, and thermal fluctuations, which
makes exact theoretical descriptions highly nontrivial. For
reviews see Refs.~\cite{RevModPhys.60.781,RevModPhys.60.1129}. 

While finite critical temperatures arise from interchain coupling,
the experimental observation of $T_c$ values much smaller than mean-field
predictions \cite{Pouget2016332} suggests that the latter is
much smaller than intrachain couplings. Above a dimensional crossover
temperature $T_\text{3D}\ll T_c$, 1D models such as the Holstein \cite{Ho59a}
and the Su-Schrieffer-Heeger (SSH) model \cite{PhysRevLett.42.1698} reviewed in
Refs.~\cite{PhysRevB.87.075149,PhysRevB.91.245147} can be used. Except for
the spinful SSH model, quantum lattice fluctuations destroy the ordered state
for a sufficiently weak electron-phonon coupling
\cite{PhysRevB.27.1680,JeZhWh99}. Beyond the critical coupling, quantum fluctuations mainly reduce the dimerization
\cite{PhysRevB.33.5141,PhysRevB.25.7789}. The ground-state properties have
been characterized in terms of correlation functions and excitation spectra
\cite{PhysRevB.87.075149,PhysRevB.91.245147}, but open questions remain concerning critical couplings and Luttinger parameters \cite{PhysRevB.92.245132}.

The numerical calculation of thermodynamic properties or spectral functions at finite
temperature, as studied experimentally
\cite{PhysRevLett.32.769,WEI1977595,PhysRevB.19.4723,0295-5075-19-6-014,PhysRevB.89.201116},
is much more difficult and limited by the large Hilbert space (for density-matrix
renormalization group methods \cite{PhysRevB.74.235118}) or the analytic continuation
(for quantum Monte Carlo methods \cite{PhysRevB.87.075149,PhysRevB.91.245147}). A few
results are available for spin-Peierls models
\cite{PhysRevB.60.12125,PhysRevB.70.214429}.  Interestingly, even the simpler
case of classical phonons has only been studied at very low
temperatures \cite{1996MPLB...10..467M,Michielsen:1997aa}. It is 
routinely used in material-specific modeling of ground-state properties, and
should provide a reliable description when the Peierls gap and/or the
temperature are large compared to the phonon frequency  \cite{Brazovskii1976}. 

Here, we systematically explore the temperature dependence of the
specific heat and the excitation spectra of spinless Holstein and SSH
models in the adiabatic limit. The latter provides the rare opportunity of
obtaining exact numerical results on large systems, including exact
high-resolution spectral functions. This allows a detailed, quantitative understanding of thermal fluctuations and a comparison to the
widely used mean-field approximation for experimentally relevant quantities
such as the specific heat and the single-particle spectral function
\cite{PhysRevLett.32.769,WEI1977595,PhysRevB.19.4723,0295-5075-19-6-014,PhysRevB.89.201116}.

The organization is as follows. In Sec.~\ref{Sec:Models} we define
the models and review their ground-state properties. The
method is described in Sec.~\ref{Sec:Method}. Results for
thermodynamic and spectral properties are discussed in Sec.~\ref{Sec:Thermo} 
and Sec.~\ref{Sec:Spec}, respectively. Section~\ref{Sec:Conclusions} contains
our conclusions, and the Appendix discusses finite-size effects.

\section{Models}
\label{Sec:Models}

We study electrons in one dimension coupled to the lattice, as described by a Hamiltonian
\begin{align}
\label{Ham}
\ham{} = \ham{\mathrm{ph}} + \ham{\mathrm{el}} \, ,
\end{align}
where $\ham{\mathrm{ph}}$ is the lattice contribution and
$\ham{\mathrm{el}}$ contains the electronic and electron-phonon parts. In general, $\ham{\mathrm{ph}}$ depends on the lattice
displacements $\Q{i}$ and momenta ${\P{i}}$. In the adiabatic limit,
the lattice is static and the displacements become classical variables $q_i$,
allowing us to replace $\ham{\mathrm{ph}}\to \hamcl{\mathrm{ph}}$ in Eq.~(\ref{Ham}). In the
following, we define the Holstein and SSH models directly in
this limit. 

The spinless Holstein model \cite{Ho59a} describes fermions
coupled to harmonic oscillators with quadratic potential
\begin{align}
\hamcl{\mathrm{ph}}
  =
  \frac{K}{2}\sum_i \q{i}^2
\end{align}
and spring constant $K$. The electronic part of the Hamiltonian is given by
\begin{align}
\label{HS_el}
\ham{\mathrm{el}}
  =
  -t \sum_i \left( \fcr{i} \fan{i+1} + \fcr{i+1} \fan{i} \right)
  + g \sum_i \q{i} \left( \fden{i} - 1/2 \right)
  \, .
\end{align}
The first term describes the nearest-neighbor hopping of spinless
fermions with amplitude $t$, where $\fcr{i}$ ($\fan{i}$) creates
(annihilates) a fermion at site $i$. In the second term, the
displacement $\q{i}$ couples to the local fermion density $\fden{i}
= \fcr{i} \fan{i}$ with coupling parameter $g$.

In the spinless SSH model \cite{PhysRevLett.42.1698}, the lattice energy
depends on the relative displacements of neighboring sites,
\begin{align}
\hamcl{\mathrm{ph}}
  =
  \frac{K}{2}\sum_i \left( \q{i+1} - \q{i} \right)^2\,.
\end{align}
The electronic part,
\begin{align}
\label{SSH_el}
\ham{\mathrm{el}}
  =
  \sum_i \left[ -t + \alpha \left( \q{i+1} - \q{i} \right)  \right]
  \left( \fcr{i} \fan{i+1} + \fcr{i+1} \fan{i} \right)
  \, ,
\end{align}
describes the modulation of the hopping amplitude by the coupling of
the lattice displacements to the bond density.

For both models, we introduce a dimensionless coupling parameter
$\lambda$ by rescaling the displacement fields. For the Holstein
model $\lambda=g^2/(4Kt)$, whereas for the SSH model
$\lambda=\alpha^2/(Kt)$. We use $t$ as the unit of energy, set the
lattice constant and $\hbar$ to one, and consider half-filling (one electron
per two sites).

At zero temperature, the exact properties of both models can
be obtained from mean-field theory \cite{PhysRevB.22.2099,PhysRevB.27.1680,PhysRevB.27.4302}.
For any $\lambda>0$, the Peierls instability leads to a dimerization of the lattice that is
captured by the ansatz $q_i = (-1)^{i} \Delta / (2g)$ for the Holstein
model and $q_i = (-1)^{i} \Delta / (8 \alpha)$ for the SSH
model. Here, $\Delta$ is the gap calculated self-consistently
from the gap equation. The lattice dimerization is
accompanied by charge-density-wave order in the Holstein model and
bond-density-wave order in the SSH model. The order has
periodicity $2\kF$, where $\kF=\pi/2$ is the Fermi momentum. Commensurability with the lattice pins the phase of the order
parameter to $\pi$ \cite{LEE1974703}, so that the ground state is twofold
degenerate under $\Delta\to-\Delta$. While exact at $T=0$, 
mean-field theory predicts a finite Peierls transition temperature $T_c$, in violation of the Mermin-Wagner theorem \cite{PhysRevLett.17.1133}.
The adiabatic limit is
expected to capture the physics of the dimerized phase  \cite{Brazovskii1976}.

While the Holstein and the SSH model both describe Peierls insulators,
important differences arise from their different symmetries. The
mean-field SSH Hamiltonian is often considered as the
simplest model of a symmetry-protected
topological band insulator \cite{PhysRevLett.89.077002},
as reviewed in Ref.~\cite{TopoBook}.
It obeys time-reversal, particle-hole, and chiral symmetry.
Explicitly, under time reversal, $\mathcal{T} \hat{c}_j
\mathcal{T}^{-1} = \hat{c}_j$ with $\mathcal{T} \im \mathcal{T}^{-1} =
-\im$, whereas for a particle-hole transformation $\mathcal{P} \hat{c}_j
\mathcal{P}^{-1} = (-1)^{j} \hat{c}^{\dagger}_j$ with $\mathcal{P} \im
\mathcal{P}^{-1} = \im$. The chiral symmetry operator is
given by $\mathcal{C} = \mathcal{T} \mathcal{P}$. These symmetries
put the SSH model into the so-called BDI class of the general classification of
symmetry-protected topological phases
\cite{PhysRevB.78.195125,Kitaev.09,1367-2630-12-6-065010} which in 1D allows for a
nontrivial topological invariant.
The two degenerate ground states of the SSH model belong to different topological sectors. The symmetry-protected zero-energy states
of the topological phase are identical to the soliton excitations at domain
walls introduced in Refs.~\cite{PhysRevLett.42.1698,PhysRevD.13.3398}. For
periodic boundaries, domain walls can only occur as soliton-antisoliton
pairs. Depending on their size, such pairs may form bound polaron states with
nonzero energy \cite{RevModPhys.60.781}. 
The Hamiltonian of the Holstein model belongs to the AI symmetry class with
broken chiral (and particle-hole) symmetry as a result of the density-displacement
coupling. The two degenerate ground states are therefore trivial and do not support
topologically protected zero-energy states at domain walls. Nevertheless,
soliton-antisoliton pairs can exist and were reported in simulations of the
quantum phonon case \cite{PhysRevB.83.115105}.
While the topological classification is strictly valid only at $T=0$,
the electronic symmetries persist for any configuration of displacements generated by thermal fluctuations. 

\section{Method}
\label{Sec:Method}

To solve the electron-phonon problem at finite temperatures, we used the
Monte Carlo method of Ref.~\cite{1996MPLB...10..467M}. In the adiabatic
limit, and using the notation of Ref.~\cite{PhysRevB.83.165203},
the partition function of Hamiltonian (\ref{Ham}) takes the form
\begin{align}
\label{partsum}
Z
  =
  \int d\q{1} \,\,\dots \int d\q{L} \,  e^{-\beta \hamcl{\mathrm{ph}}}
  Z_{\mathrm{el}}[\q{1},\dots,\q{L}]
  \, ,
\end{align}
where $Z_{\mathrm{el}} = \Tr \exp [-\beta(\ham{\mathrm{el}} - \mu
\hat{N} )]$ is the grand-canonical partition function of the
electronic subsystem, $\beta = 1/k_\text{B}T$ the inverse temperature, $\mu$
the chemical potential and $\hat{N}$ the total particle-number operator. 

For each configuration $C=\{\q{1},\dots,\q{L}\}$ of the classical displacements,
$\ham{\mathrm{el}}$ is a noninteracting Hamiltonian that can
be diagonalized exactly. The Monte Carlo method of Ref.~\cite{1996MPLB...10..467M}
samples the continuous space of displacement configurations $C$. Expectation values take the form
\begin{align}
\label{MC}
\expv{\hat{O}}
=
\sum_C W[C] \expvc{\hat{O}} 
\end{align}
with the weight of the configuration
\begin{align}
\label{weight_conf}
W[C]
=
\frac{1}{Z} e^{-\beta \hamcl{\mathrm{ph}}[C]} Z_{\mathrm{el}}[C]
\end{align}
and the corresponding value of the observable
\begin{align}
\label{obs_conf}
\expvc{\hat{O}}
=
\frac{1}{Z_{\mathrm{el}}[C]}
\Tr \left\{ e^{-\beta(\ham{\mathrm{el}}[C]-\mu \hat{N})} \hat{O}[C]
  \right\}
\, .
\end{align}
The weight $W[C]$ is always positive and can be sampled using the
Metropolis algorithm \cite{1953JChPh..21.1087M}. For each configuration,
observables are calculated from Eq.~(\ref{obs_conf}). Both quantities are obtained from a 
diagonalization of the $L \times L$ matrix representation of
$\ham{\mathrm{el}}[C]$ which dominates the computational complexity of the algorithm.

Technically, Monte Carlo simulations of Eq.~(\ref{MC}) are related to 
disorder problems at finite temperature \cite{Brazovskii1976}. For
each configuration $C$, we solve an Anderson model
\cite{PhysRev.109.1492} with either diagonal (site) disorder for the Holstein
model or off-diagonal (bond) disorder for the SSH model. In contrast to
common disorder problems, the probability distribution $W[C]$ has a
nontrivial dependence on $Z_{\mathrm{el}}[C]$. However, in the
high-temperature limit, $Z_{\mathrm{el}}[C]\approx1$ and $W[C]$
becomes a Gaussian distribution. We will revisit this analogy below.

\subsection{Sampling}

Simulations were started from random configurations which were then
updated by randomly picking a single $\q{i}$ and proposing a change
$\Delta\q{}$. $\Delta\q{}$ was drawn from a Gaussian distribution
with variance $\sigma_{q}^2$. Because at high temperatures $W[C]$ is
dominated by $\exp(-\beta\hamcl{\mathrm{ph}}[C])$, $\sigma_{q} \sim \sqrt{T}$
is a natural choice. However, at low temperatures, the distribution of
displacements evolves into a two-peak structure
\cite{doi:10.1143/JPSJ.69.2634} and $\sigma_{q} \sim \sqrt{T}$ becomes
too sharp. Therefore, for each temperature, we performed a warmup to estimate
the actual distribution of displacements. At low
temperatures, the algorithm suffers from long autocorrelation times,
which were overcome by parallel tempering
\cite{doi:10.1143/JPSJ.65.1604}. For each coupling parameter
$\lambda$, the data shown were generated from a
fixed temperature grid with at least $64$ points. A switch of
configurations at adjacent temperatures was proposed every $500$ updates.
We set $\mu = 0$ for half-filling and simulated
lattices of length $L=162$ with periodic boundary conditions.

\subsection{Observables}
\label{Sec:Obs}

In the following, we define the relevant static and dynamic observables.
For each configuration $C$, they were calculated from the
single-particle basis of $\ham{\mathrm{el}}[C]$ given by the
eigenvalues $E_{\lambda}$ and eigenvectors $\ket{\lambda}$.

The specific heat $C_V$ was calculated via
\begin{align}
\label{obs:CV}
C_V[C]
  =
  k_\text{B} \beta^2 \left[ \expvc{\ham{}^2} - \expvc{\ham{}}^2 \right]
  \,.
\end{align}
To study the ordering of the electronic subsystem, we used the static
structure factors
\begin{align}
\label{struc_stat}
S_{\alpha}(q;C)
  =  \frac{1}{L} \sum_{i,j} e^{\im q(i-j)} \expvc{\hat{O}_i^{\alpha}
  \hat{O}_j^{\alpha}}
\end{align}
as a function of transferred momentum $q$.
The subscript $\alpha=\rho$ ($\alpha=b$) denotes the charge (bond) structure factor. The corresponding operators
$\hat{O}_i^{\alpha}$ are the local charge density $\fden{i}$ and bond density $\fbond{i} = (\fcr{i}\fan{i+1} + \fcr{i+1}\fan{i})$.

Importantly, spectral functions can be calculated directly for real frequencies, without the need of numerical analytic
continuation. For the single-particle spectral function $A(k,\omega)$,
the Lehmann representation reads
\begin{align}\label{spectral_sp}
A(k,\omega;C)
  =
  \sum_{\lambda}
  \absolute{\bra{0} \fan{k} \ket{\lambda}}^2
  \delta ( \omega - E_{\lambda} ) \, .
\end{align}
From Eq.~(\ref{spectral_sp}), the density of states $N(\omega)$ was
obtained by summation over momentum $k$.
Two-particle spectra were calculated from the dynamic structure factors
\begin{align}\label{eq:twopart}
S_{\alpha}(q,\omega;C)
 &=
\Big|\sum_{\lambda} p_{\lambda}\bra{\lambda} \hat{O}_{q}^{\alpha}
  \ket{\lambda}\Big|^2 \delta(\omega)
 \\\nonumber
&+  \sum_{\lambda,\nu}
  p_{\nu}(1-p_{\lambda})
  \absolute{\bra{\lambda} \hat{O}_{q}^{\alpha}  \ket{\nu}}^2 
  \delta ( E_{\lambda} - E_{\nu} - \omega ) \, ,
\end{align}
where $p_{\lambda} = \{\exp[\beta(E_{\lambda}-\mu)]+1\}^{-1}$ is the
Fermi function and $\alpha=\rho,b$ as before.
We also consider the real part of the optical conductivity
\begin{align}\label{eq:ocdef}
\sigma(\omega;C)
  =
  \sum_{\lambda,\nu}
  \frac{p_{\nu} - p_{\lambda}}{\omega}
  \absolute{\bra{\lambda} \hat{J} \ket{\nu}}^2
  \delta ( E_{\lambda} - E_{\nu} - \omega ) \, ,
\end{align}
where $\hat{J}=\im \sum_i t_i (\fcr{i}\fan{i+1} -
\fcr{i+1}\fan{i})$ is the current operator; here $t_i = t$ for the
Holstein model and $t_i = t - \alpha (\q{i+1} - \q{i})$ for the SSH
model, respectively.

Spectral functions were measured on a discrete frequency grid. Each data
point represents the averaged spectral weight in an interval of width
$\Delta\omega$. Unless stated otherwise we used $\Delta\omega=0.04t$.

\section{Thermodynamics}
\label{Sec:Thermo}

We first discuss thermodynamic properties, focusing on the
specific heat. The latter is an integrated quantity accessible to experiments
that already captures the relevant temperature scales of the physical system.

\begin{figure}[tbp]
\centering
\includegraphics[width=\linewidth]{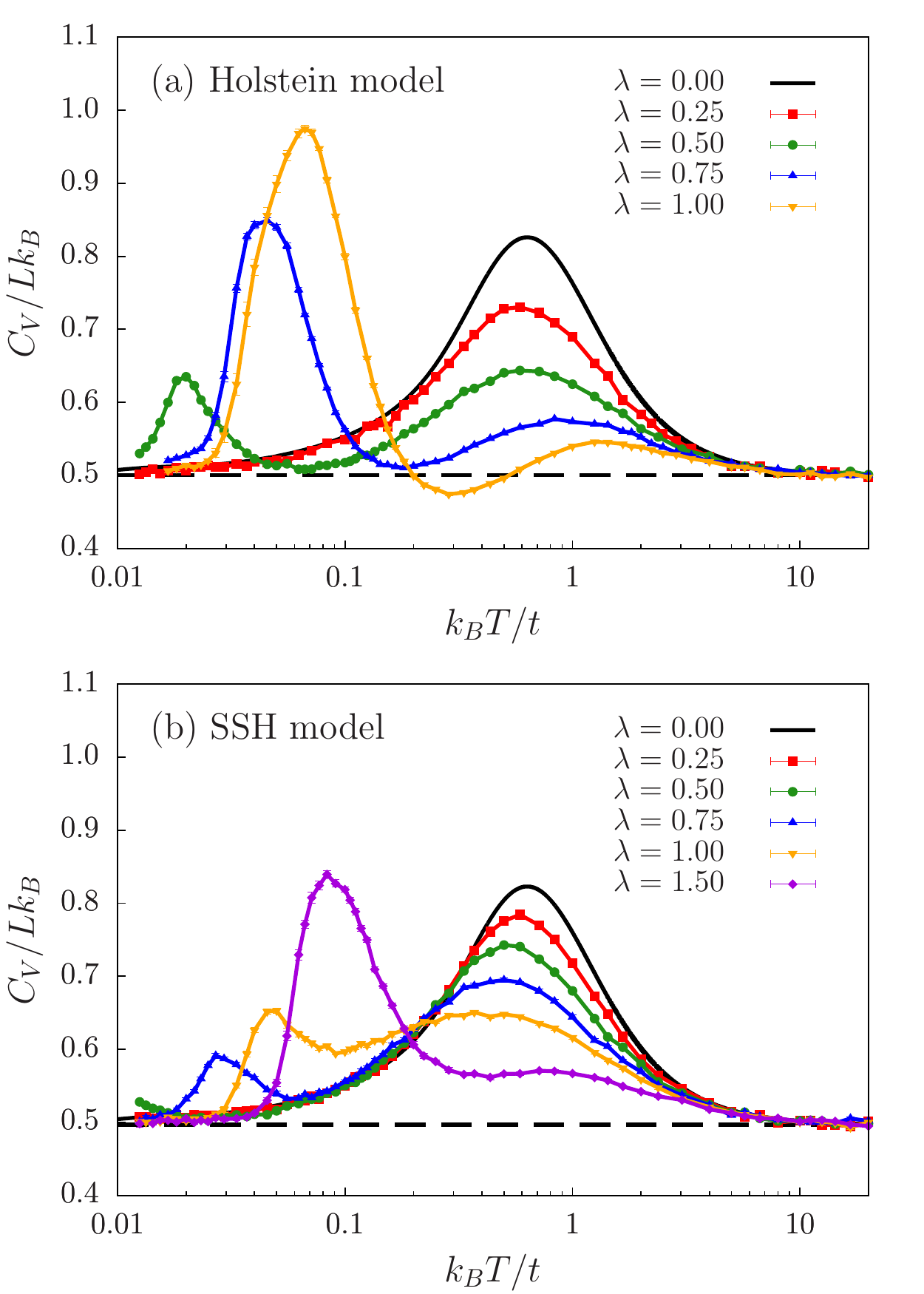}%
\caption{\label{CV}
(Color online) Specific heat per site for (a) the Holstein and (b) the SSH model with $L=162$.
The dashed lines indicate the respective free-phonon contributions.
}
\end{figure}

Figure~\ref{CV} shows the specific heat of both models as a function
of temperature and for different couplings $\lambda$. For the large lattice
size $L=162$ used, only minor finite-size effects appear (see Appendix).
Note that adjacent data points in Fig.~\ref{CV} are not statistically independent since
they were generated by parallel tempering.

At $\lambda=0$, the specific heat is the sum of contributions
from the phonons and the electrons. In the adiabatic
limit, the phonons are described by classical harmonic
oscillators. According to the equipartition theorem, each phonon mode
contributes $k_\text{B}/2$, which leads to the constant
background in Fig.~\ref{CV}. (For the SSH model, the $k=0$
mode does not contribute because the length of the chain was fixed.)
Therefore, $C_V$ does not vanish for $T \to 0$, in violation of the third law
of thermodynamics.
The electronic contribution reaches a maximum at the
coherence temperature $k_\text{B}T \approx 0.63t$ and vanishes for
$T \to 0$ and $T \to \infty$. The maximum is related to the thermal
activation of charge fluctuations across the entire band width of our
lattice model. The expected linear free-fermion contribution is
visible in the interval $0.03t < k_B T <0.1t$ (for the
system size $L=162$ used) in a different representation (not shown).

For $\lambda>0$, the electronic and phononic contributions to
$C_V$ can no longer be separated. For the Holstein
model, a small coupling $\lambda=0.25$ suppresses
$C_V$ over the whole temperature range shown in Fig.~\ref{CV}(a). With
increasing $\lambda$, the free-electron peak loses weight and shifts to higher temperatures. At $\lambda=1$ and
intermediate temperatures, the specific heat even falls below the free-phonon
contribution. For the SSH model, Fig.~\ref{CV}(b), $C_V$ is also suppressed at high
temperatures, but its maximum shifts to  slightly lower
temperatures. Moreover, $C_V$ remains almost constant at intermediate
temperatures.

For both models, an additional peak emerges in $C_V$ at low temperatures.
While for small $\lambda$ the peak cannot be observed
in the accessible temperature range, it shifts to higher temperatures
and grows with increasing $\lambda$. This feature is robust
against finite-size effects, only the downturn towards $T\to0$ where
the electronic contribution vanishes is not yet fully converged with
$L$. For a detailed finite-size analysis see the Appendix.

\begin{figure}[tbp]
\centering
\includegraphics[width=\linewidth]{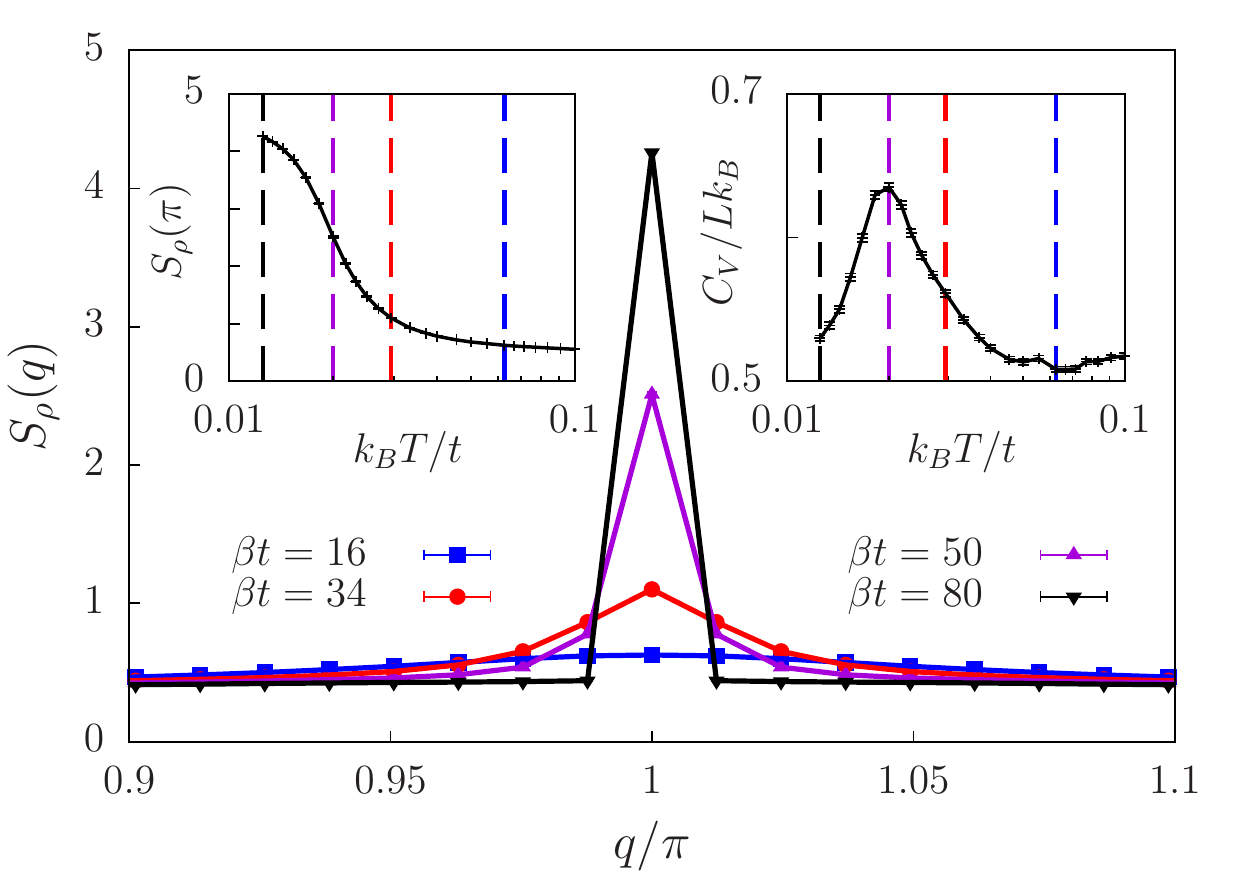}%
\caption{\label{denstrucHS}
(Color online)
Density structure factor $S_{\rho}(q)$ of the Holstein model around
the ordering vector $q=2\kF=\pi$ for selected temperatures. The full
temperature dependence of $S_{\rho}(\pi)$ is shown in the left inset.
Dashed lines mark the temperatures for which $S_{\rho}(q)$ is shown in
the main panel and match the evolution of the low-temperature peak in
$C_V$, as shown in the right inset. Here, $\lambda = 0.5$ and $L=162$. 
}
\end{figure}

The appearance of the low-temperature peak can be attributed to an
enhancement of order as temperature is decreased. Figure~\ref{denstrucHS} shows the static
density structure factor $S_{\rho}(q)$ for the Holstein model at
$\lambda=0.5$. At low temperatures, $S_{\rho}(q)$ develops a peak at
$q=2\kF=\pi$ that indicates the formation of a charge-density
wave. Simultaneously, the peak in $C_V$ arises, as shown in the two
insets of Fig.~\ref{denstrucHS}. Its maximum at $k_\text{B}T \simeq 0.02t$
corresponds with the inflection point of $S_{\rho}(\pi)$. The width of
the peak is related to the temperature range where $2\kF$ correlations
become prominent. The same behavior is expected for the SSH model and
the bond structure factor $S_b(q)$.

While true long-range order only exists at $T=0$, the position of the
low-temperature peak in $C_V$ can be regarded as a coherence scale
at which pronounced $2\kF$ correlations set in and which marks the
emergence of a clear Peierls energy gap. The thermal crossover is
described by a correlation length $\xi(T)$ \cite{Schulz1987}. While $\xi(T)\to\infty$ for
$T\to0$, corresponding to long-range order, the correlation length is finite
at $T>0$ where charge or bond correlations decay exponentially.
Similar results have been
obtained from a Ginzburg-Landau approach \cite{PhysRevB.6.3409}. While a saddle-point
approximation gives a second-order phase transition at a finite $T_c$
and a jump in $C_V$ \cite{Landau37}, Scalapino
\etal~\cite{PhysRevB.6.3409} used a functional method to treat
fluctuations in the Ginzburg-Landau fields. Thereby, they mapped
the 1D electron-phonon problem to a single quantum mechanical
anharmonic oscillator \cite{doi:10.1080/00018738200101398}. In this
approach, long-range order is destroyed at $T>0$, and
$C_V$ is continuous with a peak similar to our results. The maximum in
$C_V$ may be located well below the mean-field value for $T_c$
\cite{doi:10.1080/00018738200101398}. For the electron-phonon models
considered here, the mean-field critical temperature is an order of
magnitude larger than the peak positions in our $C_V$ data.

The results in Fig.~\ref{CV} are very similar for
the two models considered. With increasing $\lambda$, the free-electron contribution 
is suppressed and an additional low-temperature peak emerges
that can be attributed to enhanced $2\kF$ charge or bond correlations, respectively. The same
temperature scales will also be relevant for the spectral properties
discussed in Sec.~\ref{Sec:Spec}. The relation between $C_V$ and the spectral
function becomes apparent by considering the relation $C_V = \partial
E_{\mathrm{tot}} / \partial T$ and using the equation of motion
\cite{KadanoffBaym} to write the total energy as
\begin{align}
\label{EnergySpectrum}
E_{\mathrm{tot}}
  =
  \frac{N_{\mathrm{ph}}}{2\beta} +
  \sum_k \int_{-\infty}^{\infty} d\omega \,
  \frac{\omega+\epsilon_k}{2} \,
  n_F(\omega)\,
  A(k,\omega)\,.
\end{align}
Here, $N_{\mathrm{ph}}=L$ for the Holstein model and
$N_{\mathrm{ph}}=L-1$ for the SSH model. According to
Eq.~(\ref{EnergySpectrum}), $E_{\mathrm{tot}}$ can be expressed as a
sum rule of the single-particle spectrum weighted with the Fermi
function $n_F(\omega)$ and the bare dispersion $\epsilon_k =
-2t \cos k$. Thus, the specific heat measures the change of the density of
states around the Fermi energy with temperature. The decrease
of the free-electron peak in $C_V$ with increasing $\lambda$ therefore
corresponds to a reduction of spectral weight across a broad region of
energies and temperature, whereas the sharp low-temperature peak
signals a sudden change in the single-particle spectrum. In particular, we
will show that the emergence of the low-temperature peak is related to the
Peierls gap.

\section{Spectral properties}
\label{Sec:Spec}

In this section, we investigate how the temperature-driven suppression of
$2\kF$ charge or bond order manifests itself in the single-particle and
two-particle spectral functions
[Eqs.~(\ref{spectral_sp})--(\ref{eq:ocdef})]. While at $T=0$ the
spectral functions can be calculated exactly using mean-field
theory, finite temperatures require numerical simulations.

\subsection{Holstein model}

For the Holstein model, the electron-phonon coupling is chosen as
$\lambda=0.5$, for which the mean-field gap $\Delta \approx 0.68t$ and the
interesting temperature scale set by the corresponding peak in $C_V$ is well accessible.

\subsubsection{Temperature dependence of the density of states}

\begin{figure}[tbp]
\centering
\includegraphics[width=\linewidth]{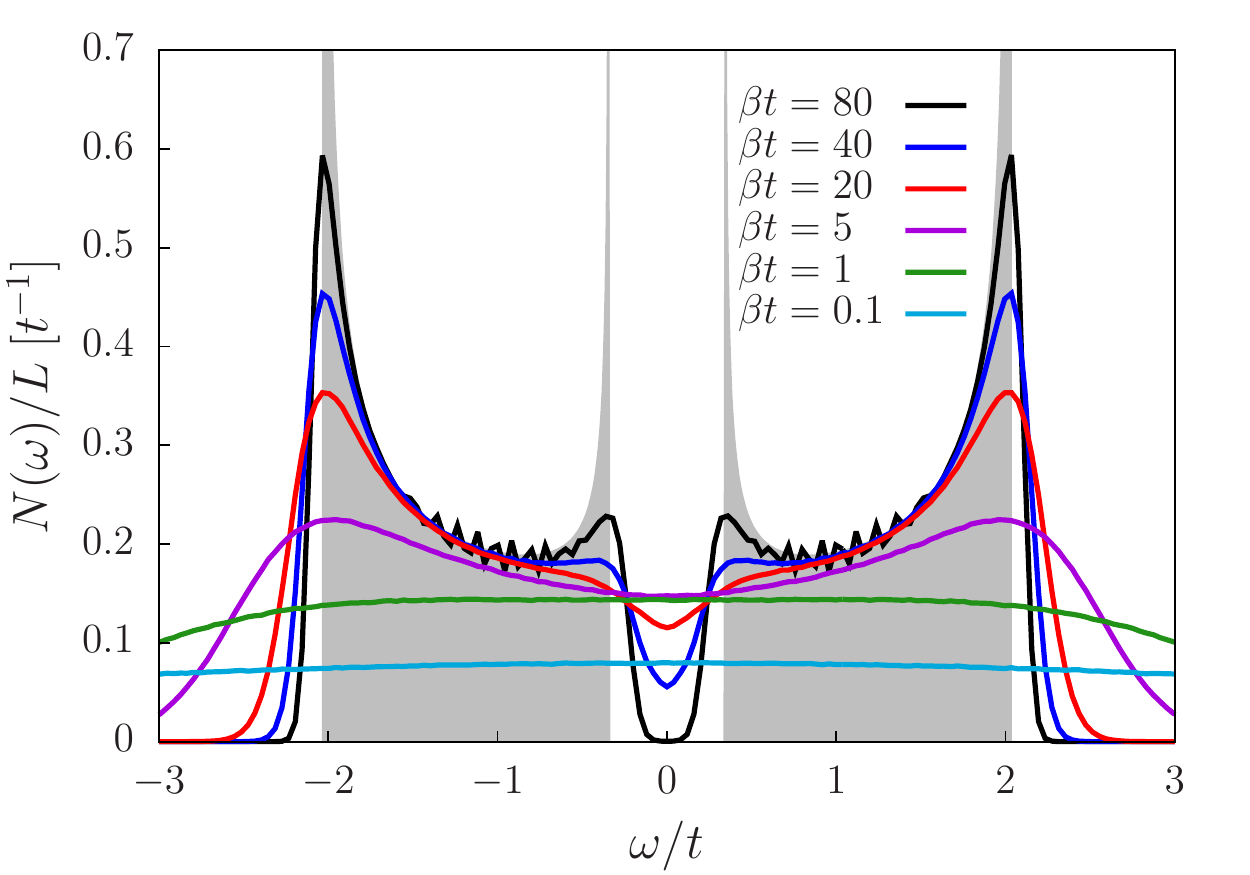}%
\caption{\label{DOS_HS}
(Color online)
Density of states of the Holstein model for $\lambda =0.5$ and $L=162$. The
filled curve corresponds to the $T=0$ mean-field result~(\ref{Eq:DOS_HS}).
}
\end{figure}

\begin{figure*}[ht]
\includegraphics[width=\linewidth]{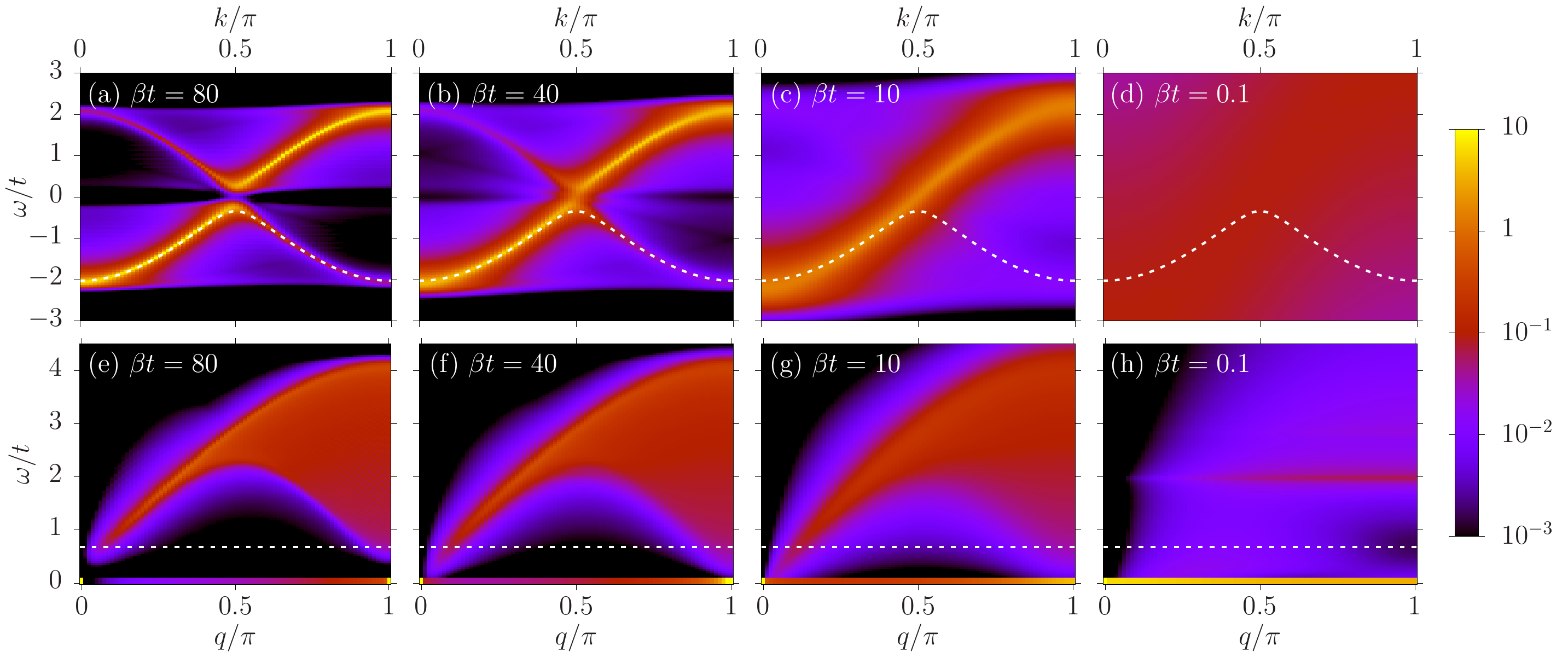}%
\caption{\label{spectral_HS}
(Color online)
(a)--(d) Single-particle spectral function $A(k,\omega)$ and (e)--(h) dynamic
density structure factor $S_{\rho}(q,\omega)$ of the Holstein model for
$\lambda = 0.5$ and $L=162$.  Dashed lines correspond to the $T=0$ mean-field dispersion
and gap, respectively. For better visibility, the $\omega=0$
contributions to $S_{\rho}(q,\omega)$ are shown as a bar of width $0.1t$ in (e)--(h).
}
\end{figure*}

We begin with the density of states plotted in Fig.~\ref{DOS_HS}.
The filled curve shows the exact mean-field result at $T=0$ which in the
thermodynamic limit is given by
\begin{align}
\label{Eq:DOS_HS}
N(\omega)
  =
  \frac{L}{\pi}
  \frac{\absolute{\omega}}
  {\sqrt{\omega^2-\left(\frac{\Delta}{2}\right)^2}
   \sqrt{\left(2t\right)^2+\left(\frac{\Delta}{2}\right)^2-\omega^2}}
\end{align}
for $\Delta/2 < \absolute{\omega} < \sqrt{(2t)^2+(\Delta/2)^2}$, and zero else.
Hence, at the mean-field level, the electron-phonon
interaction opens a gap $\Delta$ at the Fermi level and the shifts the upper edge of the
band to higher energies. At the band edges, square-root
singularities appear. 

Thermal fluctuations lead to a broadening of the band edges and the 
singularities become finite peaks. At the lowest temperature considered in our
simulation, $\beta t = 80$, $N(\omega)$ is still close to the result
at $T=0$, but spectral weight enters the mean-field gap
exponentially. The fine structure visible in the middle of the bands is a
finite-size effect and is partly smeared due to the use 
of a frequency grid with spacing $\Delta\omega$. With increasing
temperature, the peak at the lower edge of the spectrum is strongly
suppressed. At the same time, the $T=0$ gap is filled in and has disappeared at $\beta t =5$. At even higher temperatures, also
the peak at the upper edge is entirely washed out. The weight is shifted to
higher frequencies and the spectrum flattens completely.

The temperature of the gap closing in Fig.~\ref{DOS_HS}
coincides with the position of the low-temperature peak in $C_V$ and the
suppression of $2\kF$ correlations in $S_\rho(q)$ in Fig.~\ref{denstrucHS}. According to
Eq.~(\ref{EnergySpectrum}), the change of $N(\omega)$ near the Fermi
level is largest at the coherence scale $k_\text{B}T \approx 0.02t$ where $C_V$ has its
maximum. Therefore, the peak in $C_V$ directly signals the formation
of the gap. Its temperature scale is considerably lower than the
mean-field gap $\Delta/2 \approx 0.34 t$ or the critical temperature
$k_\text{B}T_c \approx 0.2 t$, similar to the reduction of the
transition temperature due to 1D fluctuations in
Refs.~\cite{PhysRevB.6.3409,PhysRevLett.31.462}.

\subsubsection{Momentum dependence of the spectral functions}

The single-particle spectrum $A(k,\omega)$ and the dynamic density structure
factor $S_{\rho}(q,\omega)$ are shown in Fig.~\ref{spectral_HS}. The
temperatures were chosen to capture the interesting regions defined by
the results for $C_V$ in Fig.~\ref{CV}.

For $\beta t = 80$ [Fig.~\ref{spectral_HS}(a)], $A(k,\omega)$ closely
follows the mean-field dispersion indicated by the dashed line. The imbalance of spectral weight between the original cosine
dispersion and the shadow bands is characteristic for systems with competing
periodicities and only disappears for $\lambda\to\infty$ \cite{Voit501}.
Due to the finite temperature, the peaks in $A(k,\omega)$ are broadened and
their positions deviate slightly from the mean-field dispersion at the
band edges. There are additional features of minor weight that
disperse from the edges of the original cosine band forming a
continuum of excitations.

With increasing temperature [Fig.~\ref{spectral_HS}(b)], the broadening
becomes larger and the shadow bands less pronounced. Inside
the mean-field gap, two dispersing bands appear with dominant weight
around $k_\text{F}=\pi/2$ (see also Sec.~\ref{Sec:HS_gap_closing}).
At $\beta t = 10$ [Fig.~\ref{spectral_HS}(c)], the gap and the shadow bands have
disappeared completely, and the locus of spectral weight follows the cosine
dispersion of the noninteracting system. Further increasing the
temperature only leads to a broadening of the spectrum until it
becomes washed out completely, see Fig.~\ref{spectral_HS}(d).

Figures~\ref{spectral_HS}(e)--(h) show the dynamic density structure
factor $S_{\rho}(q,\omega)$ at the same temperatures. At $\beta t =
80$, $S_{\rho}(q,\omega)$ exhibits a particle-hole continuum but
with a gap comparable to the mean-field gap (dashed line). Moreover,
there is a sharp central (Bragg) peak at $q=2\kF=\pi$ associated with
charge-density-wave order. At higher temperature
[Figs.~\ref{spectral_HS}(f)--(g)], the edges of the particle-hole continuum
diffuse, the gap is filled in, and the central peak becomes a Lorentzian of width
$\xi^{-1}(T)$ in momentum space (cf. Fig.~\ref{denstrucHS}) where $\xi(T)$
is the correlation length introduced at the beginning of Sec.~\ref{Sec:Thermo}.
In the high-temperature limit [Fig.~\ref{spectral_HS}(h)]
the particle-hole continuum is washed out completely, and
$S_{\rho}(q,\omega)$ contains (i) a spatially localized (\ie,
$q$-independent) zero-energy Einstein phonon mode, and
(ii) an additional mode at $\omega=2t$ related to the strong onsite disorder
generated for the fermions by the lattice fluctuations (see Sec.~\ref{Sec:HighTemp}).

\subsubsection{Closing of the single-particle gap}
\label{Sec:HS_gap_closing}

The closing of the single-particle gap in Fig.~\ref{spectral_HS} is the
result of two effects. First, a spatially homogeneous renormalization of the
$T=0$ mean-field order parameter. Second, thermally induced defects in the
lattice dimerization with energies below the band gap.

\begin{figure}[tbp]
\hspace{-0.4cm}
\includegraphics[width=\linewidth]{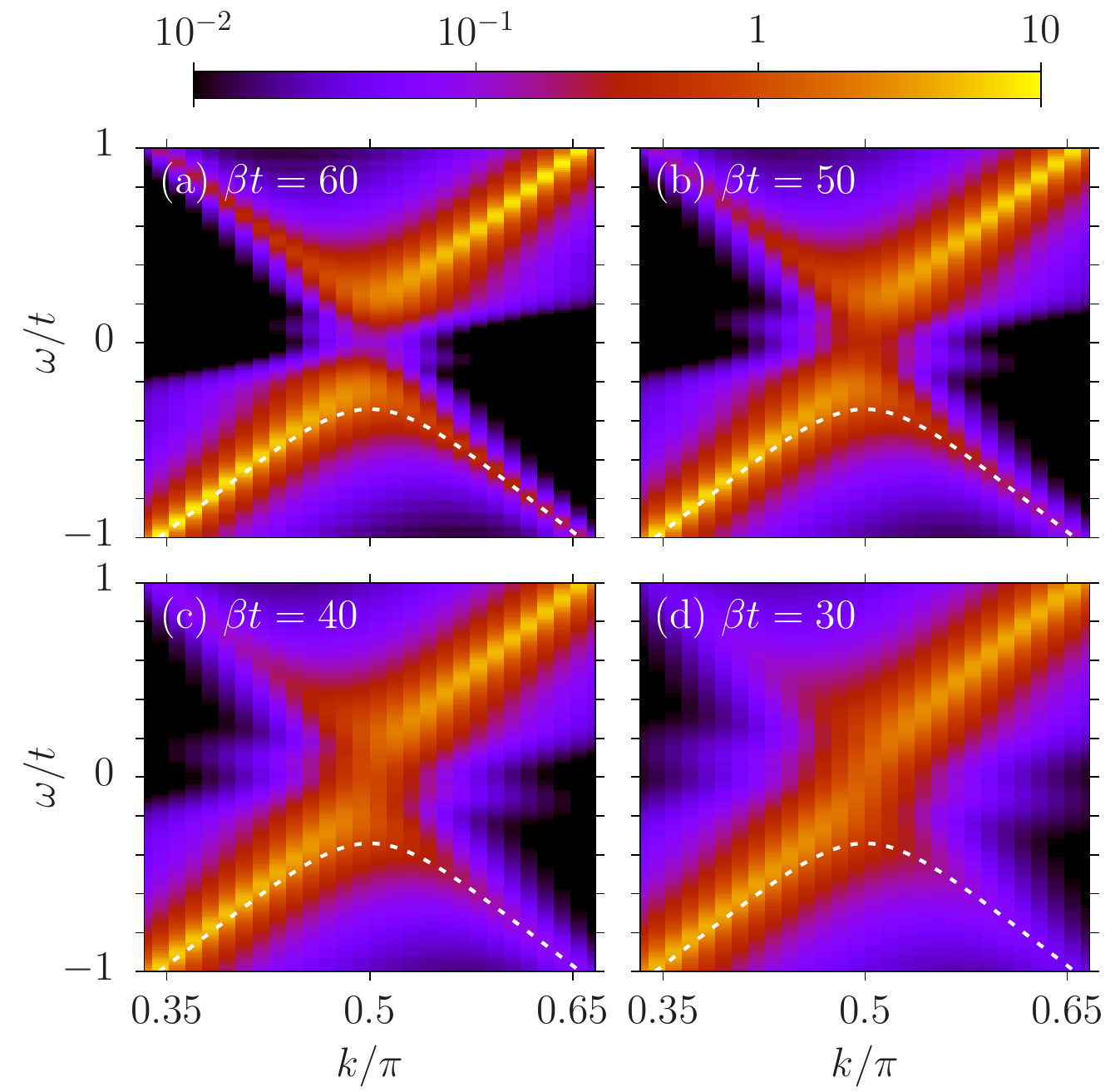}%
\caption{\label{spectral_HS_gap}
(Color online)
Close-up of the single-particle spectral function $A(k,\omega)$
around $\kF=\pi/2$ for the same parameters as in Fig.~\ref{spectral_HS}.
 The dashed lines correspond to the mean-field dispersion
at $T=0$. Here, we used a discretization $\Delta\omega=0.01t$.
}
\end{figure}

A closeup of the thermally induced low-energy excitations is shown
in Fig.~\ref{spectral_HS_gap}. For $\beta t = 60$
[Fig.~\ref{spectral_HS_gap}(a)], we see a band above (below) the
mean-field main band for $k<\kF$ ($k>\kF$), as well as a weaker 
band below (above) the mean-field shadow band for $k<\kF$ ($k>\kF$) 
that extends only over a small range of $k$ around $\kF$. Both features
merge with the mean-field bands near $\kF$. With increasing temperature, 
the additional excitations gain spectral weight (especially close to
$\kF$) and the feature following the shadow bands extends over a large
$k$-range. Eventually, the gap is filled in and the linear dispersion near
$\kF$ is restored, cf. Figs.~\ref{spectral_HS}(c) and \ref{spectral_HS}(d).

At low temperatures [Fig.~\ref{spectral_HS}(a)], the spectral function has a
close resemblance with that of the spinless Holstein model with quantum
phonons \cite{PhysRevB.83.115105}. The latter exhibits dispersive excitations
with energy smaller than the mean-field gap that have been interpreted as
polaron excitations. While quantum fluctuations reduce the minimal energy for
polaron excitations \cite{PhysRevB.83.115105}, the latter coincides with the mean-field gap in the
classical case [Fig.~\ref{spectral_HS}(a)].

\subsubsection{Optical conductivity}

\begin{figure}[tbp]
\centering
\includegraphics[width=\linewidth]{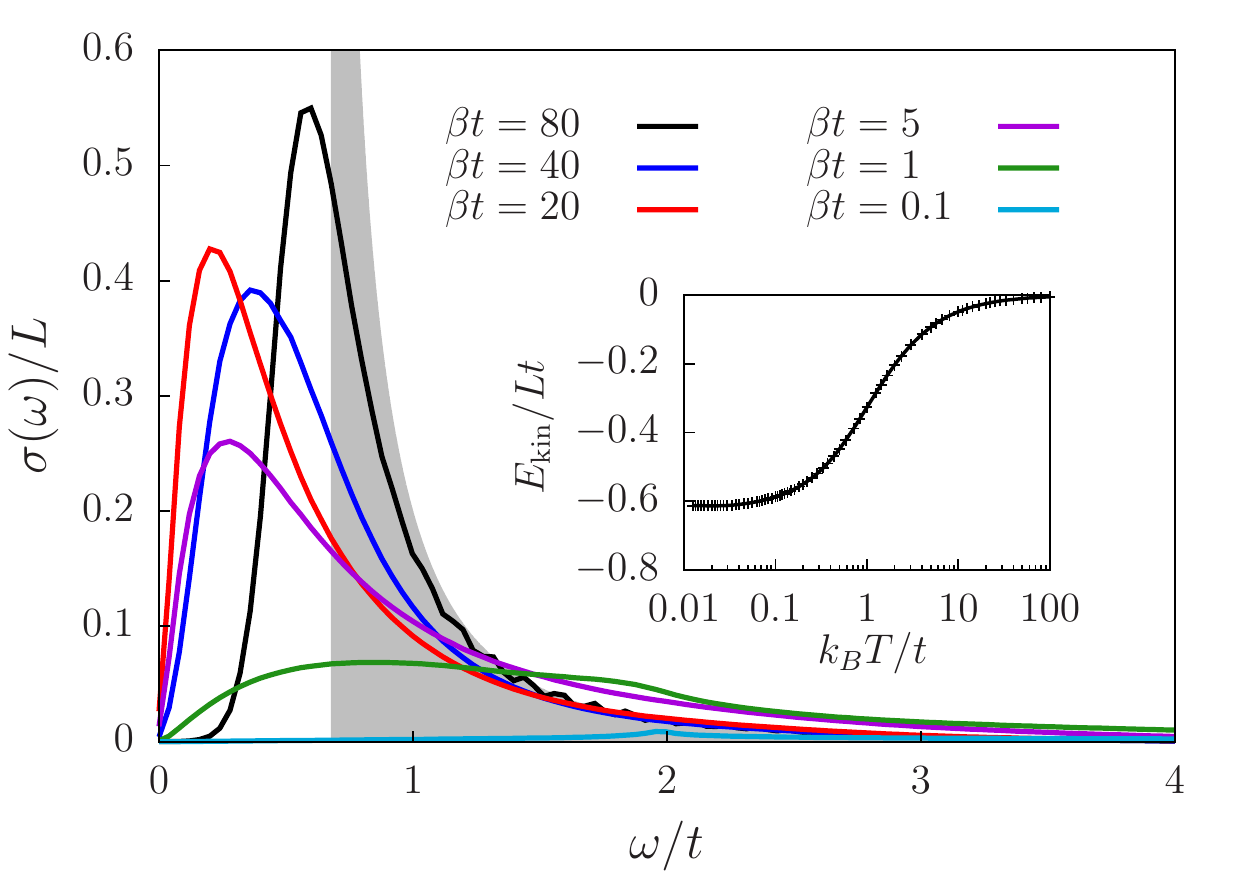}%
\caption{\label{opt_cond_HS}
(Color online)
Optical conductivity of the Holstein model for $\lambda =0.5$ and
$L=162$. The filled curve is the $T=0$ mean-field result~(\ref{eq:oc_mf}). The inset
shows the kinetic energy of the electrons as a function of
temperature. It is related to $\sigma(\omega)$ by the sum rule given
in Eq.~(\ref{HS_sum_rule}).
}
\end{figure}

Finally, we consider the optical conductivity $\sigma(\omega)$ in Fig.~\ref{opt_cond_HS}.
At $T=0$, mean-field theory gives
\begin{align}\label{eq:oc_mf}
\sigma(\omega)
  =
  \frac{L \Delta^2}{4\pi \omega^2}
  \sqrt{\frac{(4t)^2+\Delta^2-\omega^2}{\omega^2-\Delta^2}}
\end{align}
for $\Delta < \absolute{\omega} < \sqrt{(4t)^2+\Delta^2}$. The filled
curve in Fig.~\ref{opt_cond_HS} clearly shows the square-root
singularity at the lower edge $\omega=\Delta$. In
contrast to the density of states, there is no singularity at the
upper edge where $\sigma(\omega)=0$. At $\beta t = 80$,
the lower edge of $\sigma(\omega)$ has already broadened
significantly.
As a function of temperature, we first observe a decrease of the optical gap
due to the suppression of charge order. While this shift is
qualitatively captured by a temperature-dependent mean-field gap $\Delta(T)$,
the latter does not account for the nontrivial broadening due to fluctuations. Although the
single-particle gap is filled in at high temperatures, there is no Drude peak.
The absence of the latter, and the shift of the peak in $\sigma(\omega)$ back to
larger frequencies for $\beta t \lesssim 20$, can be attributed to the onset
of incoherence. In contrast, in the mean-field charge-density-wave
approximation, $\Delta=0$ at $T>T_c$ so that the electrons can move coherently. 
At even higher temperatures, the strong lattice fluctuations act as
essentially random disorder. A characteristic peak emerges at 
$\omega=2t$ that becomes more pronounced as temperature increases further.
The relation to a disorder problem will be discussed in more detail in
Sec.~\ref{Sec:HighTemp}.

The integrated optical conductivity is related to the kinetic energy via
the f-sum rule \cite{PhysRevB.56.4484}
\begin{align}
\label{HS_sum_rule}
\int_0^{\infty} d\omega \, \sigma(\omega)
  =
  - \frac{\pi}{2} E_{\mathrm{kin}}
\,.
\end{align}
The results for $E_\text{kin}$ in the inset of Fig.~\ref{opt_cond_HS} reveal
that up to $\beta t \approx 20$ spectral weight is merely redistributed,
whereas it is significantly reduced at higher temperatures and vanishes for $T\to\infty$.

\subsection{SSH model}

The spectral properties of the SSH model are in many aspects
similar to the Holstein model, and we therefore focus on the differences.
To facilitate a comparison with the results for the Holstein model
we take $\lambda=0.75$ for which the mean-field gap $\Delta \approx 0.76t$.

\subsubsection{Temperature dependence of the density of states}

\begin{figure}[tbp]
\centering
\includegraphics[width=\linewidth]{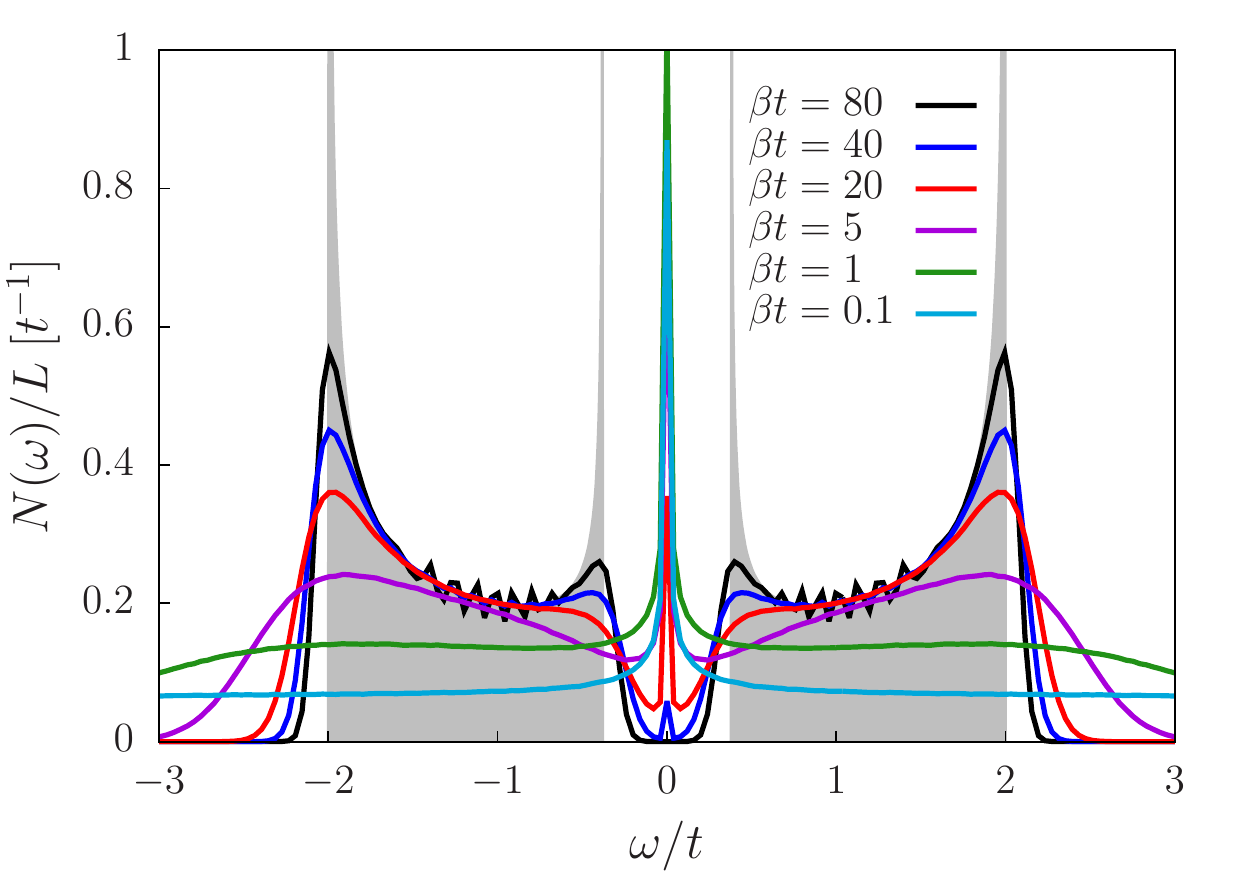}%
\caption{\label{DOS_SSH}
(Color online)
Density of states of the SSH model for $\lambda=0.75$ and $L=162$. The
filled curve corresponds to the $T=0$ mean-field result~(\ref{Eq:DOS_SSH}).
}
\end{figure}

Figure~\ref{DOS_SSH} shows the density of states, including the 
$T=0$ mean-field result given by
\begin{align}
\label{Eq:DOS_SSH}
N(\omega)
  =
  \frac{L}{\pi}
  \frac{\absolute{\omega}}
  {\sqrt{\omega^2-\left(\frac{\Delta}{2}\right)^2}
   \sqrt{(2t)^2-\omega^2}}
\end{align}
for $\Delta/2 < \absolute{\omega} < 2t$, and zero otherwise.
Equation~(\ref{Eq:DOS_SSH}) has the same form as Eq.~(\ref{Eq:DOS_HS}), but
the upper edge of the spectrum remains at $\omega=2t$ independent of
$\lambda$. The temperature dependence of the mean-field bands, \ie, the
broadening of the singularities and the closing of the gap, is similar
to the Holstein model. However, there is an additional peak at
$\omega=0$ that grows and broadens with increasing temperature. It survives
even at the highest temperature considered where the rest of the spectrum has
been completely washed out by thermal fluctuations. As discussed below,
the peak is related to topologically protected midgap states of the SSH Hamiltonian. 

\subsubsection{Momentum dependence of the spectral functions}

\begin{figure*}[tbp]
\includegraphics[width=\linewidth]{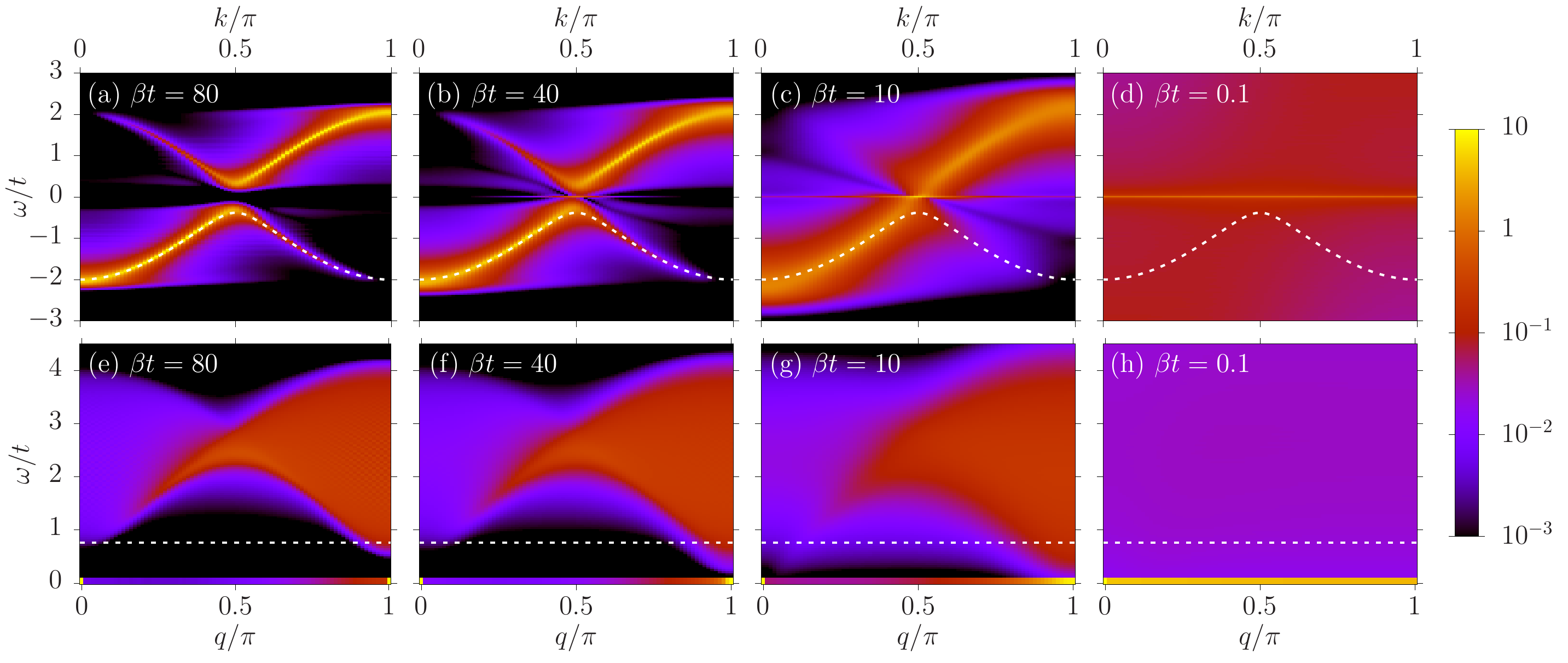}%
\caption{\label{spectral_SSH}
(Color online)
(a)--(d) Single-particle spectral function $A(k,\omega)$ and (e)--(h) dynamic
bond structure factor $S_b(q,\omega)$ of the SSH model for $\lambda = 0.75$
and $L=162$. The dashed lines correspond to the $T=0$ mean-field dispersion
and gap, respectively.
For better visibility, the $\omega=0$
contributions to $S_{b}(q,\omega)$ are shown as a bar of width $0.1t$ in (e)--(h).
}
\end{figure*}

The single-particle spectral function $A(k,\omega)$ shown in
Figs.~\ref{spectral_SSH}(a)--(d) is again very similar to the Holstein
model, except for the zero-energy peak. The latter is absent at  $\beta t = 80$
[Fig.~\ref{spectral_SSH}(a)], where the spectrum closely follows the
mean-field dispersion. It first emerges at $\beta t\simeq
40$ when the gap starts to be filled in by thermal excitations
[Fig.~\ref{spectral_SSH}(b)]. 
At $\beta t =10$ [Fig.~\ref{spectral_SSH}(c)], the mean-field gap is filled
in but signatures of the shadow bands remain. More noticeably, 
the zero-energy peak is well visible for all $k$ with maximal spectral
weight at $\kF$. Finally, increasing the temperature further to $\beta t =
0.1$ completely smears out the spectrum except for the $\omega=0$ peak
[Fig.~\ref{spectral_SSH}(d)]; in this regime, the spectral weight of the peak
becomes independent of $k$.

The corresponding results for the dynamic bond structure factor are shown in
Figs.~\ref{spectral_SSH}(e)--(h). At the lowest
temperature considered [Fig.~\ref{spectral_SSH}(e)], it has a
continuum of excitations above the mean-field gap and
zero-energy peaks at $q=0$ and $q=2\kF=\pi$. The evolution with
  temperature is similar to Fig.~\ref{spectral_HS}. In particular, the gap
  is filled in and the Lorentzian central peak widens due to the 
  decrease of $\xi(T)$. In the high-temperature limit
[Fig.~\ref{spectral_SSH}(h)], sharp excitations exist only at $\omega=0$.

\subsubsection{Localization of the zero-energy mode}

\begin{figure}[tbp]
\includegraphics[width=\linewidth]{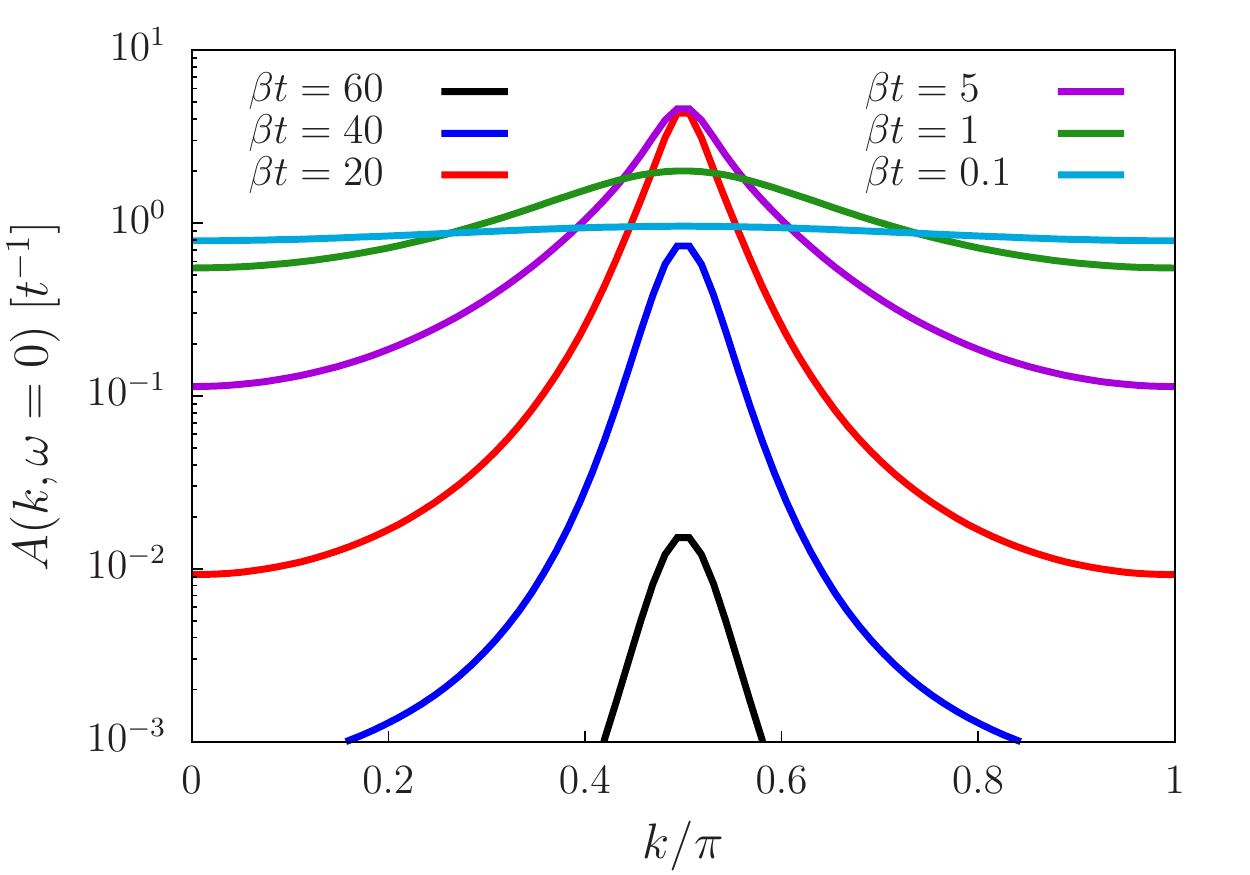}%
\caption{\label{soliton_SSH}
(Color online)
Temperature dependence of the zero-energy peak in $A(k,\omega)$ as a
function of momentum $k$ for the SSH model for $\lambda = 0.75$ and
$L=162$. The spectrum was averaged over an interval $\Delta\omega=0.04t$
around $\omega=0$. Hence, the extent of the peak in frequency
is not captured. 
}
\end{figure}

We attribute the zero-energy mode in the single-particle spectrum to 
soliton states at thermally generated domain walls between different
lattice dimerizations \cite{PhysRevLett.42.1698,PhysRevB.22.2099}. 
We can estimate the spatial extent of these states from their 
momentum dependence, which is shown in Fig.~\ref{soliton_SSH}.
At low temperatures, the shape of the peak hardly
changes, only its spectral weight becomes larger. A comparison with
the analytic result for the soliton wave function \cite{PhysRevB.22.2099},
$\phi_0(n) \sim \sech(n/l) \cos(\pi n/2)$, gives a localization
length of $l\approx5$ in units of the lattice spacing, in agreement
with  Ref.~\cite{PhysRevB.22.2099}. As the
temperature exceeds $\beta t = 20$, the peak in Fig.~\ref{soliton_SSH}
broadens in $k$-space and the localization length becomes
smaller. In the high-temperature limit, the zero-energy state becomes
completely localized. Although the picture of domain walls between
ordered regions breaks down when the single-particle gap closes, the
zero mode persists at higher temperatures [Fig.~\ref{spectral_SSH}(d)] 
where it can be understood as a disorder effect, see Sec.~\ref{Sec:HighTemp}.

\subsubsection{Optical conductivity}

\begin{figure}[tbp]
\centering
\includegraphics[width=\linewidth]{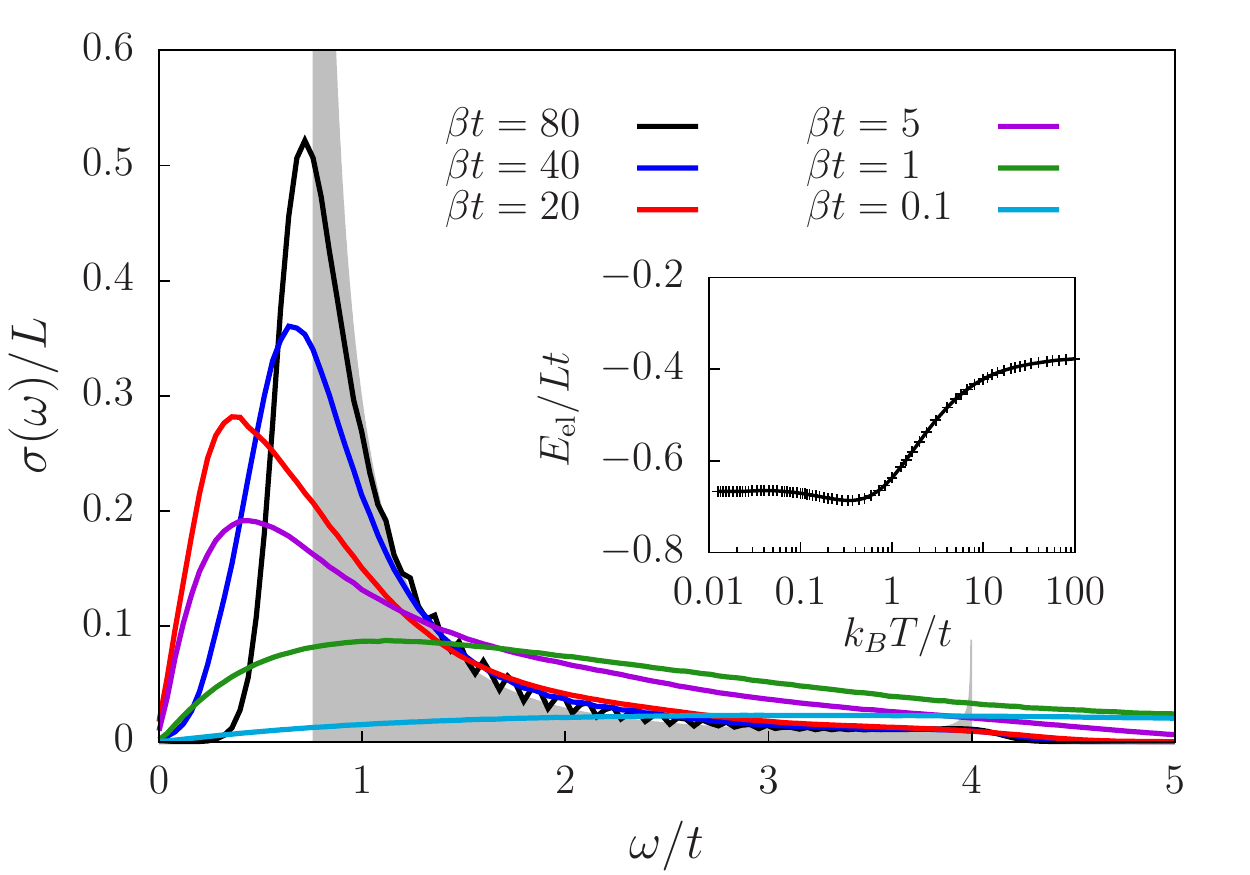}%
\caption{\label{opt_cond_SSH}
(Color online)
Optical conductivity of the SSH model for $\lambda =0.75$ and
$L=162$. The filled curve is the $T=0$ mean-field result~(\ref{eq:oc_mf_ssh}). The inset
shows the energy of the electronic subsystem as a function of
temperature. It is related to $\sigma(\omega)$ by the sum rule given
in Eq.~(\ref{SSH_sum_rule}). 
}
\end{figure}
The optical conductivity $\sigma(\omega)$ is shown in
Fig.~\ref{opt_cond_SSH}. At $T=0$, the mean-field result is given by
\begin{align}\label{eq:oc_mf_ssh}
\sigma(\omega)
  =
  \frac{4 L \Delta^2 t^2}{\pi \omega^2}
  \frac{1}{\sqrt{\omega^2-\Delta^2}}
  \frac{1}{\sqrt{(4t)^2-\omega^2}}
\end{align}
for $\Delta < \absolute{\omega} < 4t$, otherwise it is zero. Compared
to the Holstein model, it has an additional square-root singularity
at the upper edge of the spectrum. However, its integrated weight is
too small to be visible even at the lowest temperature considered. The lower
edge first broadens and then also shifts to lower frequencies.
Similar to the Holstein model, up to $\beta t \approx 10$ spectral weight is
only redistributed, as visible from the
inset of Fig.~\ref{opt_cond_SSH}. The integrated spectrum is related
to the energy of the electronic subsystem via the sum rule
\cite{PhysRevB.56.4484}
\begin{align}
\label{SSH_sum_rule}
\int_0^{\infty} d\omega \, \sigma(\omega)
  =
  - \frac{\pi}{2} E_{\mathrm{el}}
  \,.
\end{align}
In contrast to the Holstein model, the sum rule also includes
the interaction energy of electrons and phonons. Because of this
contribution, the integrated weight slightly increases between $\beta
t \approx 10$ and $\beta t \approx 3$.
Further increasing the temperature leads to a reduction of spectral
weight at small $\omega$ and a substantial enhancement of the tail at
large $\omega$. In contrast to the Holstein model, the integrated
weight does not vanish for $T\to\infty$.

\subsection{Relation to disorder problems}
\label{Sec:HighTemp}

At high temperatures, the essentially random lattice distortions act as 
disorder for the electrons \cite{Brazovskii1976}, corresponding to
site disorder for the Holstein model, and bond disorder for the SSH model. The probability distribution $W[C]$
[Eq.~(\ref{weight_conf})] becomes a Gaussian and the disorder strength scales
as $\sqrt{\lambda T}$. The connection to disordered noninteracting models
explains some of the spectral features observed above.

For the Holstein model, the strong onsite disorder leads to two
distinct peaks in the two-particle spectra [Figs.~\ref{spectral_HS}(h)
and \ref{opt_cond_HS}], one at $\omega=0$ in $S_{\rho}(q,\omega)$,
and another at $\omega=2t$ both in $S_{\rho}(q,\omega)$ and
$\sigma(\omega)$. The zero-energy peak in $S_{\rho}(q,\omega)$ does
not show any $q$ dependence, whereas the peak at $\omega=2t$ is
strongest around $q=\pi$, but vanishes at $q=0$. The latter signature
also appears in $\sigma(\omega)$, where it has already been observed
for the $t-V$ model at strong disorder \cite{PhysRevB.82.161106} and
the Holstein polaron in the adiabatic regime
\cite{PhysRevB.72.104304}. This signature becomes even sharper as
temperature is increased further. In Ref.~\cite{PhysRevB.72.104304},
the resonance at $\omega=2t$ has been explained from an effective
two-site model, where the bonding and antibonding eigenstates of the
electron perfectly overlap with the current operator. In the same
way, $\hat{n}_q$ connects the different-parity states at $q=\pi$, whereas
the overlap is zero at $q=0$. 

For the SSH model, only the zero-energy peak appears in the
high-temperature limit of $S_b(q,\omega)$. Moreover, an excitation with
$\omega=0$ is visible in the single-particle spectrum and persists
for $T\to\infty$. Such a peak has previously been observed for the SSH polaron
\cite{PhysRevB.83.165203} and explained as a disorder effect
\cite{PhysRev.92.1331,PhysRevB.13.4597,PhysRevB.24.5698,0022-3719-10-6-003}. For the
tight-binding model, any finite off-diagonal disorder leads to a
zero-energy peak in the density of states that becomes larger and
broadens as the disorder strength increases
\cite{PhysRevB.13.4597}. The appearance of the peak is related to the
chiral symmetry of the SSH Hamiltonian. The latter is broken by
onsite disorder, and the zero mode disappears accordingly
\cite{PhysRevB.13.4597}. Moreover, no zero mode exists for the Holstein
model for which chiral symmetry is broken already at the mean-field level. While we have so far interpreted the
zero-energy excitations at low temperatures in terms of topologically
protected soliton states located at domain walls, such states can also
be  induced by off-diagonal disorder acting on the dimerized ground state
\cite{0295-5075-38-9-687,PhysRevLett.82.988,PhysRevB.60.15488,PhysRevB.61.12496}. 

At low temperatures, the broadening of the spectral functions
can be considered as a disorder effect, including the tail of the optical
conductivity extending into the mean-field gap. For the Holstein model, it is
related to the weak pinning of a charge-density wave by onsite
disorder \cite{PhysRevB.17.535}. For the SSH model, similar results
were also obtained from the fluctuating gap model, where order
parameter fluctuations are modeled as off-diagonal disorder
\cite{PhysRevLett.71.4015,PhysRevB.60.15488,PhysRevB.61.12496,PhysRevLett.87.126402,0295-5075-65-1-068}.

\section{Conclusions}
\label{Sec:Conclusions}

We presented exact numerical results for the thermodynamic and spectral
properties of Peierls insulators within the framework of spinless Holstein
and SSH models in the adiabatic limit. In this limit, a dimerized Peierls
state with long-range charge and lattice order exists for any nonzero
electron-phonon coupling at zero temperature, as described by mean-field theory.
Using a Monte Carlo method to sample the classical phonons, we investigated 
the impact of thermal fluctuations on the specific heat, the single-particle
spectrum, the dynamic density and bond structure factors, as well as the
optical conductivity. In contrast to the quantum case, we were able to
calculate spectra without using analytic continuation.

Thermal fluctuations destroy the mean-field long-range order, and give rise
to a characteristic low-temperature peak in the specific heat. While there is
no finite-temperature phase transition in the 1D case considered,
the peak occurs at the temperature scale at which the Peierls gap is filled in
by thermal fluctuations. A similar peak has previously been observed in theory and experiment. 
Thermally excited solitons manifest themselves as in-gap excitations. In the Holstein
model, the latter appear symmetrically around the Fermi level only at low temperatures,
whereas in the SSH model they are pinned to zero energy and persist for all
temperatures due to the chiral symmetry of the Hamiltonian.
The filling of the Peierls gap by thermal excitations is also reflected in 
the two-particle excitation spectra and the optical
conductivity. For the latter, we observe a nontrivial interplay of
  enhanced low-frequency transport due to a thermally reduced Peierls gap and
  a suppression due to the onset of incoherence. Finally,
at high temperatures, the spectral features of the electron-phonon models
are related to those of models with site or bond disorder.

The results obtained here in the adiabatic limit provide exact benchmarks.
In contrast to the widely used mean-field approximation, all thermal fluctuations (\ie, amplitude fluctuations of the
order parameter and solitons) are taken into account and finite-size effects
are negligible at all but the lowest temperatures. Our findings at low
temperatures complement previous work on the case with quantum phonons. In
particular, the spectral functions reveal many of the characteristic features
observed for quantum Holstein and SSH models \cite{PhysRevB.83.115105,PhysRevB.91.245147}. The
adiabatic approximation is expected to quantitatively capture the physics of
the quantum case for temperatures large compared to the Peierls gap. Finally,
the present findings provide a platform for the thermodynamics
of quantum phonon models, for which finite-size effects and uncertainties
related to the analytic continuation pose significant challenges to simulations.

\begin{acknowledgments}
The authors gratefully acknowledge the computing time granted by the John
von Neumann Institute for Computing (NIC) and provided on the supercomputer
JURECA \cite{Juelich} at J\"ulich Supercomputing Centre (JSC), as well as financial support
from the Deutsche Forschungsgemeinschaft (DFG) Grant Nos. AS120/10-1
and Ho~4489/3-1 (FOR 1807). We further thank J. Hofmann for helpful discussions.
\end{acknowledgments}

\appendix*
\section{Finite-size analysis of the specific heat}

\begin{figure*}[tbp]
\centering
\includegraphics[width=0.95\linewidth]{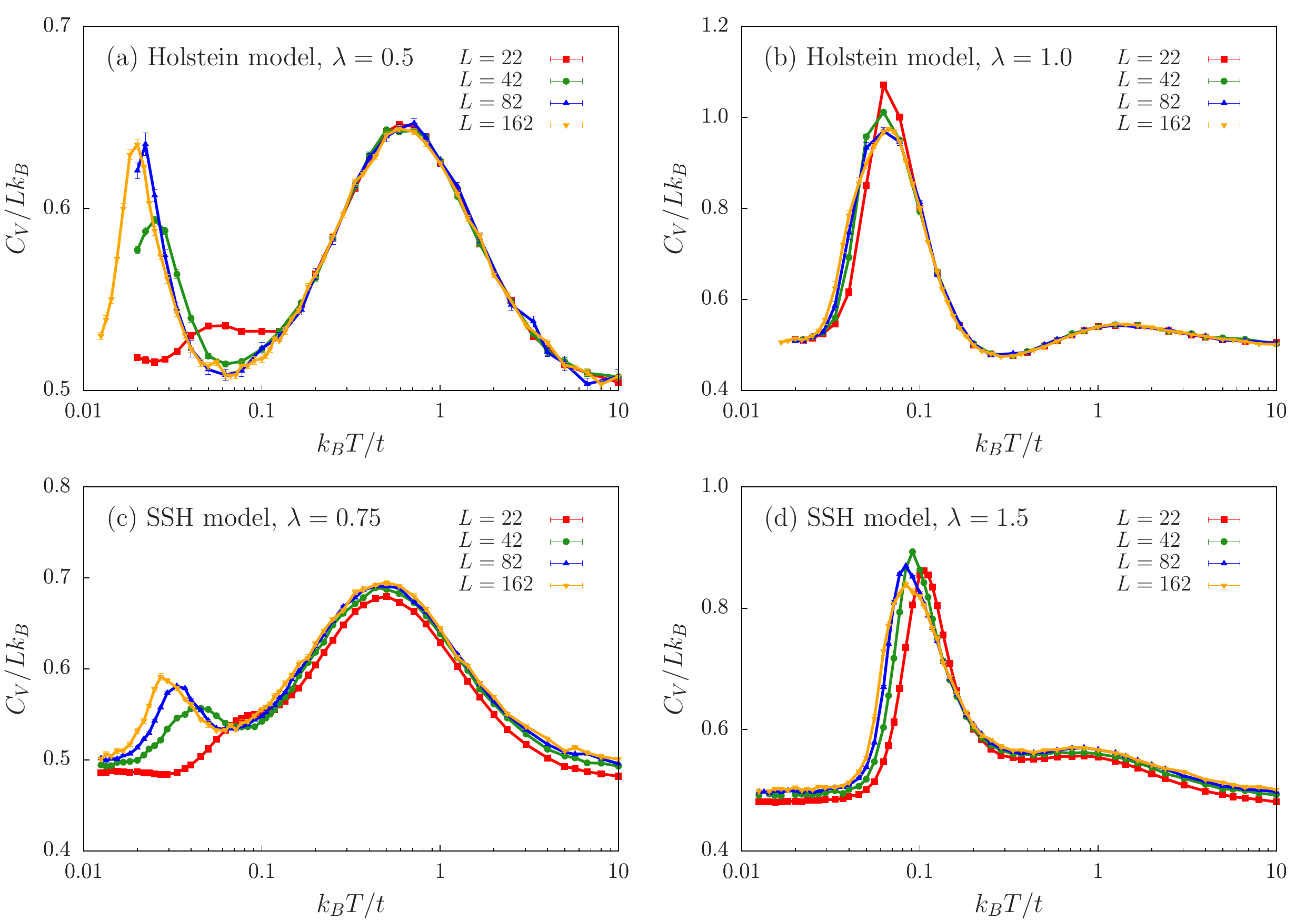}%
\caption{\label{FS}
(Color online)
Specific heat of the Holstein model
[(a), (b)] and the SSH model [(c), (d)] for different system sizes $L$.
}
\end{figure*}

In Sec.~\ref{Sec:Thermo} we discussed the low-temperature behavior of $C_V$,
and observed the appearance of a peak  related to the ordering of the
lattice. A reliable analysis also requires a study of finite-size effects.
Therefore, we present in Fig.~\ref{FS}  $C_V$ as a function of temperature for
different system sizes ranging from $L=22$ to $L=162$, and for two values of the
electron-phonon coupling.

Figure~\ref{FS}(a) shows data for the Holstein model with $\lambda=0.5$. For
temperatures $k_\text{B} T > 0.1t$, $C_V$ has already converged at the smallest $L$
considered, whereas for lower temperatures a clear dependence on the lattice
size is visible. Between $L=22$ and
$L=82$, both the position of the low-temperature peak and its height
change substantially. The upturn to its maximum is only converged for the two largest lattice sizes.
 At $\lambda=1$ [Fig.~\ref{FS}(b)],
the peak appears at higher temperatures and its upturn is already
converged for $L=22$. While the height of the maximum has converged
for $L=82$, the subsequent downturn to the lowest temperatures measured still
changes from $L=82$ to $L=162$. Note that error bars are large in this
temperature regime and adjacent data points are not independent due to
the use of parallel tempering.

For the SSH model, finite-size effects  on $C_V$ are also visible at high
temperatures [Fig.~\ref{FS}(c) and~\ref{FS}(d)]. However, these effects are simply
related to the fact that only $L-1$ phonon modes contribute to $C_V$ because the
length of the chain is fixed and the $k=0$ mode drops out of the Hamiltonian.
The finite-size effects at low temperatures are slightly larger than for the
Holstein model. For $\lambda=0.75$ [Fig.~\ref{FS}(c)], the peak position and height still change up to
$L=162$. Compared to the finite-size convergence in the Holstein model
at $\lambda=0.5$ [Fig.~\ref{FS}(a)], we believe that the upturn at
$L=162$ is converged. For $\lambda=1.5$ [Fig.~\ref{FS}(d)] it
is indeed converged, but the subsequent downturn again shows
finite-size effects.

The above analysis suggests that except for the downturn at the lowest
temperatures considered, the $C_V$ data shown in Fig.~\ref{CV} have converged
with respect to $L$. The finite-size effects on $C_V$ may also be consulted in order to
estimate finite-size effects on the spectral functions.


%

\end{document}